\documentclass{article}
\usepackage{spconf,amsmath,graphicx,lipsum,tikz,adjustbox,booktabs,multirow,makecell,arydshln,url}
\usepackage{xcolor}
\usetikzlibrary{shapes.geometric}
\usepackage{xspace}
\usepackage{flushend}
\usepackage{epstopdf}
\usepackage[sort]{cite}

\title{Blind Reverberation Time Estimation in Dynamic Acoustic Conditions}

\name{Philipp G\"{o}tz$^{1}$, Cagdas Tuna$^{2}$, Andreas Walther$^{2}$, Emanu\"{e}l A. P. Habets$^{1}$\thanks{$^\dag$International Audio Laboratories Erlangen is a joint institution of the Friedrich-Alexander-University Erlangen-N\"{u}rnberg (FAU) and Fraunhofer IIS.
\newline Corresponding author: philipp.goetz@audiolabs-erlangen.de.}}
\address{$^1$International Audio Laboratories Erlangen$^\dag$, Germany.
\\ $^2$Fraunhofer Institute for Integrated Circuits IIS, Erlangen, Germany.}
\pgfdeclarelayer{figurelayer}
\pgfdeclarelayer{annotlayer}
\pgfsetlayers{figurelayer,annotlayer,main}
\tikzstyle{none}=[inner sep=0pt]
\usetikzlibrary{decorations.markings}
\begin{document}

\ninept
\newcommand{\tsixty}{$\mathrm{RT}_{60}$\xspace}
\newcommand{\ts}{\textsuperscript}
\newcommand{\mbf}[1]{\mathbf{#1}}

\maketitle
\begin{abstract}
The estimation of reverberation time from real-world signals plays a central role in a wide range of applications. In many scenarios, acoustic conditions change over time which in turn requires the estimate to be updated continuously. Previously proposed methods involving deep neural networks were mostly designed and tested under the assumption of static acoustic conditions. In this work, we show that these approaches can perform poorly in dynamically evolving acoustic environments. Motivated by a recent trend towards data-centric approaches in machine learning, we propose a novel way of generating training data and demonstrate, using an existing deep neural network architecture, the considerable improvement in the ability to follow temporal changes in reverberation time.
\end{abstract}

\setlength{\tabcolsep}{1.2pt}

\section{Introduction}
\label{sec:intro}
The time it takes the acoustic energy in a steady-state reverberant sound field to decay by $60\,\mathrm{dB}$, referred to as reverberation time \tsixty in seconds, is a parameter of central importance in the description of many acoustic environments. It is defined by the process of propagation, reflection, and absorption of acoustic waves throughout an enclosure and across its boundaries and depends on geometric and acoustic properties of the volume or space it was measured in. Though in principle not limited to, it is mainly used as a measure for architectural spaces (e.g.~rooms, halls, staircases).

In theory, \tsixty is position independent within an enclosed volume. However, especially at wavelengths comparable to the dimensions of the enclosure, variations are often observed in real-world measurements and require averaging across multiple transmission paths throughout the volume \cite{iso3382}. Another critical assumption of measuring the reverberation time -- either based on the direct measurement of the energy decay curve or via the integrated acoustic impulse response (AIR) -- is constant acoustic conditions and, traditionally, the use of dedicated measurement signals. Once conditions change, e.g., when a room door is opened or a sound source or receiver moves to a different position, a renewed \tsixty measurement is necessary. There are scenarios in which it is desired to continuously estimate the reverberation time from real-world signals and require the possibility to adapt to changes in acoustic conditions. Such applications pose the challenge of continuously extracting patterns of temporally decaying acoustic energy from signals of great variety and estimating the decay rate of the dynamic acoustic environments in which they were acquired. 

With a wide range of potential applications in areas such as acoustic scene analysis \cite{stowell2015detection}, speech enhancement \cite{benesty2011speech}, extended reality (VR/AR) \cite{remaggi2019perceived}, acoustic environment classification \cite{eaton2016estimation} or audio forensics \cite{malik2010audio}, the blind estimation of \tsixty from real-world signals has seen increased interest in the past. Prior to leveraging the pattern recognition and non-linear approximation capabilities of deep neural networks (DNNs) in the recent past, traditional statistical approaches employing signal decay rate distributions and maximum likelihood estimation were used to estimate the reverberation time from noisy signals blindly \cite{ratnam2003blind,lollmann2008estimation,6637629,lollmann2010improved,wen2008blind}. While many of these approaches produced very useful results, the non-linear approximation capabilities of DNNs pushed the state-of-the-art even further. Convolutional neural networks (CNN) were used to estimate \tsixty and Speech Transmission Index (STI) from spectro-temporal representations of noisy, reverberant speech \cite{gamper2018blind,duangpummet2021blind}. CNNs were extended by a recurrent layer to form convolutional recurrent neural networks (CRNN) that exploit sequential dependencies in the data \cite{callens2020joint,perez2019machine} and further improve estimation accuracy \cite{deng2020online}. The focus of other studies was on auditory aspects of reverberation, with attempts to match artificial reverberation with perceptual characteristics of a real recording \cite{sarroff2020blind} and the blind estimation of AIRs from reverberant speech \cite{steinmetz2021filtered}.

For approaches that involve DNNs for the estimation of \tsixty from reverberant speech, a method of generating realistic training data is essential. In recent studies \cite{gamper2018blind,deng2020online}, the reverberant speech was generated by convolving short segments of anechoic speech with measured or simulated AIRs with the assumption of static acoustic conditions, meaning that the acoustic transmission path between source and receiver captured in the AIR remained constant over time. However, real-world sound fields are often subject to temporally changing conditions, and while the solution proposed by \cite{deng2020online} is targeted towards online \tsixty estimation, the authors did not investigate the performance in dynamic acoustic conditions. The results presented in this work show that the system's performance is limited under such conditions, at least using their proposed training scheme. In this contribution, we aim to overcome this limitation and hence improve the estimation accuracy in dynamically changing acoustic environments. Motivated by a recent shift from a model-centric towards a data-centric approach to various problems in machine learning, we propose a novel way of generating training data and investigate its effect on the ability of a DNN to follow temporal changes in acoustic conditions.

The remainder of the paper is organized as follows. In Section~\ref{sec:model}, we review the CRNN architecture that is used in this study, Section~\ref{sec:data_generation} covers the different data generation methods that are proposed. In Section~\ref{sec:eval}, the estimation performance resulting from the different training methods is evaluated, Section~\ref{sec:conclusion} concludes this paper.

\section{Model Review}
\label{sec:model}
As this contribution focuses on dynamic reverberant sound fields, we choose a network architecture suitable to provide short-time estimates of the reverberation time and adopt the CRNN proposed in \cite{deng2020online}. In the following paragraphs, the input data format, the model architecture and the relationship between temporal resolutions of the model input and output are discussed.

\subsection{Model input}
\label{ssec:model_input}
Using a gammatone filterbank \cite{patterson1987efficient} of $21$ bands, with the center frequency of the lowest band set to $100~\mathrm{Hz}$, the reverberant speech signal sampled at a rate of $16\,\mathrm{kHz}$ is transformed into a time-frequency representation used in the training, validation, and testing process. Each gammatone spectrum is approximated by linear transformation of the short-time Fourier transform of the input signal \cite{gold2011speech}, gammatone spectrograms are computed with a window length and hop length of $t_w = 4\,\mathrm{ms}$ and $t_h = 2\,\mathrm{ms}$ (frame overlap of $50\,\%$), respectively.

\subsection{Model architecture}
\label{ssec:arch}
In the initial CNN section of the model, time-frequency features are extracted from the input gammatone spectrogram across six convolutional layers. Each convolution is followed by a rectified linear unit function (ReLU) and batch normalization across the five convolutional channels. The encoder structure is followed by dropout regularization that helps to avoid overfitting and a reorganization of the dimensions before the information flows through a recurrent layer (LSTM) with a cell state size and a hidden state size of $20$ neurons each. Finally, after downsampling along the frequency dimension by a factor of two using Max Pooling, a fully connected layer followed by a ReLU function yields the estimate for each time step. The model has a total of $5611$ trainable parameters, hyper-parameter optimization \cite{akiba2019optuna} was carried out to determine learning rate, batch size, and dropout probability. For further details, we refer the reader to \cite{deng2020online}.

\subsection{Receptive field}
\label{ssec:receptive_field}
It is crucial to investigate the temporal relationship between model input and output with a potential application in real-time scenarios. The CNN encoder structure at the beginning of the model contains strided convolutions across six layers that result in a significant amount of temporal compression of the input data. The total receptive field $r_0$ of this encoder structure may be computed by \cite{araujo2019computing}:
\begin{equation}
    r_0 = \sum_{l=1}^L\left((k_l - 1)\prod_{i=1}^{l-1}s_i\right) + 1,
    \label{eqn:recep}
\end{equation}
where $L$ is the number of layers that define the receptive field, $k_l$ and $s_i$ are the kernel size and stride of the $l^{th}$ and $i^{th}$ layer, respectively. Hence, a signal length of $t_s=t_h(r_0-1)+t_w=208\,\mathrm{ms}$ is taken into account for each short-time estimate of \tsixty, according to the parametrization of the CNN section proposed in \cite{deng2020online} ($r_0=103$). Following the stride specifications of the model and the hop length $t_h$ of the gammatone spectrogram computation, a new estimate is computed every $64\,\mathrm{ms}$. Such a temporal resolution allows tracking \tsixty in scenarios in which sudden changes are expected. In comparison, the previously proposed CNN-based model by Gamper and Tashev \cite{gamper2018blind} produces a single estimate for a signal length of four seconds.

\section{Data generation}
\label{sec:data_generation}

\subsection{Reverberant data}
As outlined in Sec.~\ref{ssec:model_input}, the reverberant speech used in this contribution is transformed into a time-frequency representation using a gammatone filterbank. To accelerate the training process and to achieve convergence, each gammatone spectrogram is standardized \cite{lecun2012efficient}, i.e. each sample is made zero-mean and scaled such that its standard deviation is equal to one. Furthermore, spectral whitening is applied to exclude non-linear magnitude distributions from the features learned by the neural network. In accordance with \cite{gamper2018blind} and \cite{deng2020online}, the ground truth \tsixty used during regression is obtained using the method proposed by Karjalainen et al. \cite{antsalo2001estimation}.

\begin{figure}[t]
     \centering
     \includegraphics[width=\columnwidth,trim={0 0.7cm 0 0}]{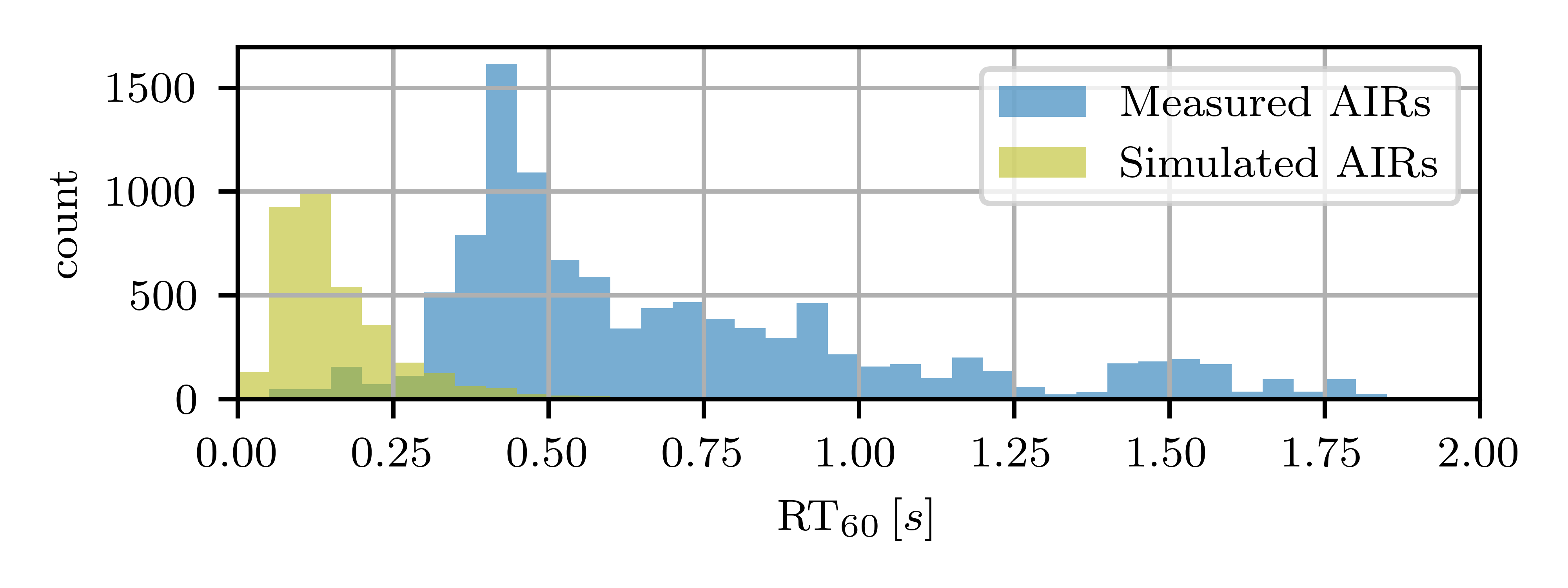}
     \caption{Distribution of \tsixty in measured and simulated AIRs.}
     \label{fig:rt60_hist}
 \end{figure}

Three anechoic speech data sets were used to generate the data in this study: LibriSpeech ASR Corpus \cite{panayotov2015librispeech}, TIMIT Acoustic-Phonetic Continuous Speech Corpus \cite{zue1990speech} and PAVOQUE Text-to-speech Corpus \cite{schroder2011open}. Measured AIRs were taken from the ACE challenge data set \cite{eaton2016estimation}, the IKS Aachen Impulse Response database \cite{jeub2009binaural}, the OpenAir database \cite{murphy2010openair} and the EchoThief database \cite{echothief}. AIRs with a reverberation time exceeding two seconds were excluded, leaving a total of $930$ measured responses with \tsixty~ranging between $0.09$~$\mathrm{s}$ and $1.95$~$\mathrm{s}$. In addition to measured data, the image method \cite{allen1979image} was used to simulate $2000$ AIRs in randomly generated enclosures with \tsixty ranging between $0.01$~$\mathrm{s}$ and $0.83$~$\mathrm{s}$. The entire data was split into training and validation data by $80\%$ and $20\%$, test data for the evaluation of all four models were generated separately, as outlined in Sec.~\ref{ssec:eval_data}. The ratio between measured and simulated AIRs in training and validation data sets was $4:1$. The distribution of \tsixty in the training data is shown in Fig.~\ref{fig:rt60_hist}.

\subsection{Static and dynamic training data}
\label{ssec:static_dynamic_data}
\begin{figure}[!t]
\begin{tikzpicture}[]
    \begin{pgfonlayer}{figurelayer}
        \node [align=left] (0) at (-3,0) {\includegraphics[width=0.99\columnwidth]{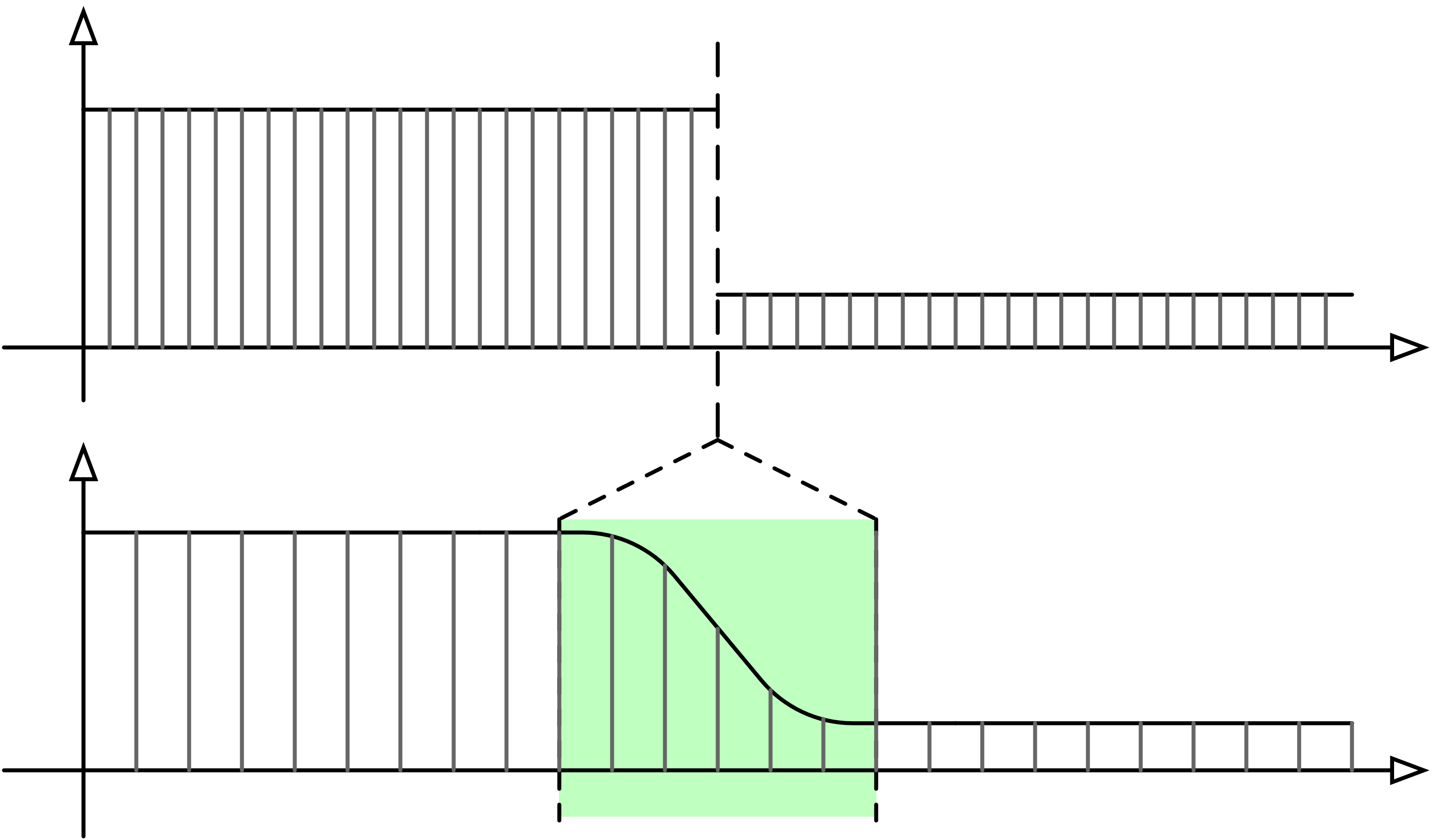}};
    \end{pgfonlayer}
    \begin{pgfonlayer}{annotlayer}
        \node [align=center] (0) at (0.2,0.22) {\textit{input time}};
        \node [align=center] (0) at (0.2,-2.3) {\textit{output time}};
        \node [align=center,rotate=90] (0) at (-7,-1) {\tsixty};
        \node [align=center,rotate=90] (0) at (-7,1.4) {\tsixty};
        \node [align=left] (0) at (-1.5,1.3) {$\leftarrow$ Change in\\acoustic conditions};
    \end{pgfonlayer}
\end{tikzpicture}
\caption{Illustration of the ground truth interpolation: At the top, the ground truth at the input time scale is shown where the dashed line marks the moment of change in acoustic conditions. The bottom plot shows the ground truth at the output time scale. The green area indicates the segment of output frames, in which the moment of transition is located within the input frames. The smooth transition is a result of the six consecutive convolutional layers in the model.}
\label{fig:gti}
\end{figure}
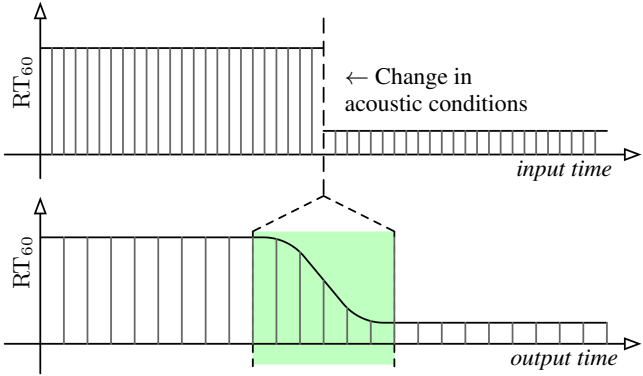

Samples of fixed duration are generated that represent reverberant speech under both static and dynamic acoustic conditions. In the static case, a single AIR is used to reverberate the anechoic speech sample, whereas in the dynamic case, two different AIRs are used, with a change occurring after a predefined duration. We investigate four different methods of generating reverberant speech samples, which are used to train, validate and test the model:
\begin{itemize}
    \item{\emph{Static conditions (4 sec.)}: Segments of anechoic speech with a duration of four seconds, convolved with a single AIR. This method was used in \cite{gamper2018blind} and \cite{deng2020online} and serves as a baseline.}
    \item{\emph{Static conditions (2 sec.)}: Segments of anechoic speech with a duration of two seconds, convolved with a single AIR. The aim is to prevent the model from learning temporal dependencies of more than two seconds.}
    \item{\emph{Dynamic conditions (deterministic)}: Samples of reverberant speech with a total duration of four seconds are generated, the AIR used to reverberate the anechoic speech is exchanged for a different one after two seconds.}
    \item{\emph{Dynamic conditions (random)}: Samples of reverberant speech with a total duration of four seconds are generated, the AIR is switched at a random moment chosen between $0.8$ and $3.2$ seconds. This method is used to rule out the possibility that the model expects the change in conditions always after two seconds, as could be an effect of the deterministic dynamic training condition.}
\end{itemize}
An efficient way of increasing training data is by recombining anechoic speech segments and AIRs to generate new, essentially unseen data. In this study, four-fold recombination was used, resulting in approximately $5252\,\mathrm{hrs.}$ of reverberant speech for training, validation, and test data for each of the four conditions. The loss function used to determine the weight update (RMSprop, adaptive learning rate optimization algorithm \cite{tieleman2012lecture}) is the mean squared error (MSE) across all time steps in each mini-batch. The change in acoustic conditions occurs between neighboring time frames in each input sample. As a result of this, and due to the temporal compression of the convolutional layers of the model, the ground truth vector is smoothed following the kernel size and stride specifications along the temporal dimension of the CNN encoder of the model. An illustrative example is shown in Fig.~\ref{fig:gti}.

\subsection{Example}
\begin{figure}[!t]
\begin{tikzpicture}[]
    \begin{pgfonlayer}{figurelayer}
        \node [align=left] (0) at (-3,0) {\includegraphics[width=\columnwidth,trim={0.0cm 0.7cm 0.0cm 0.0cm}]{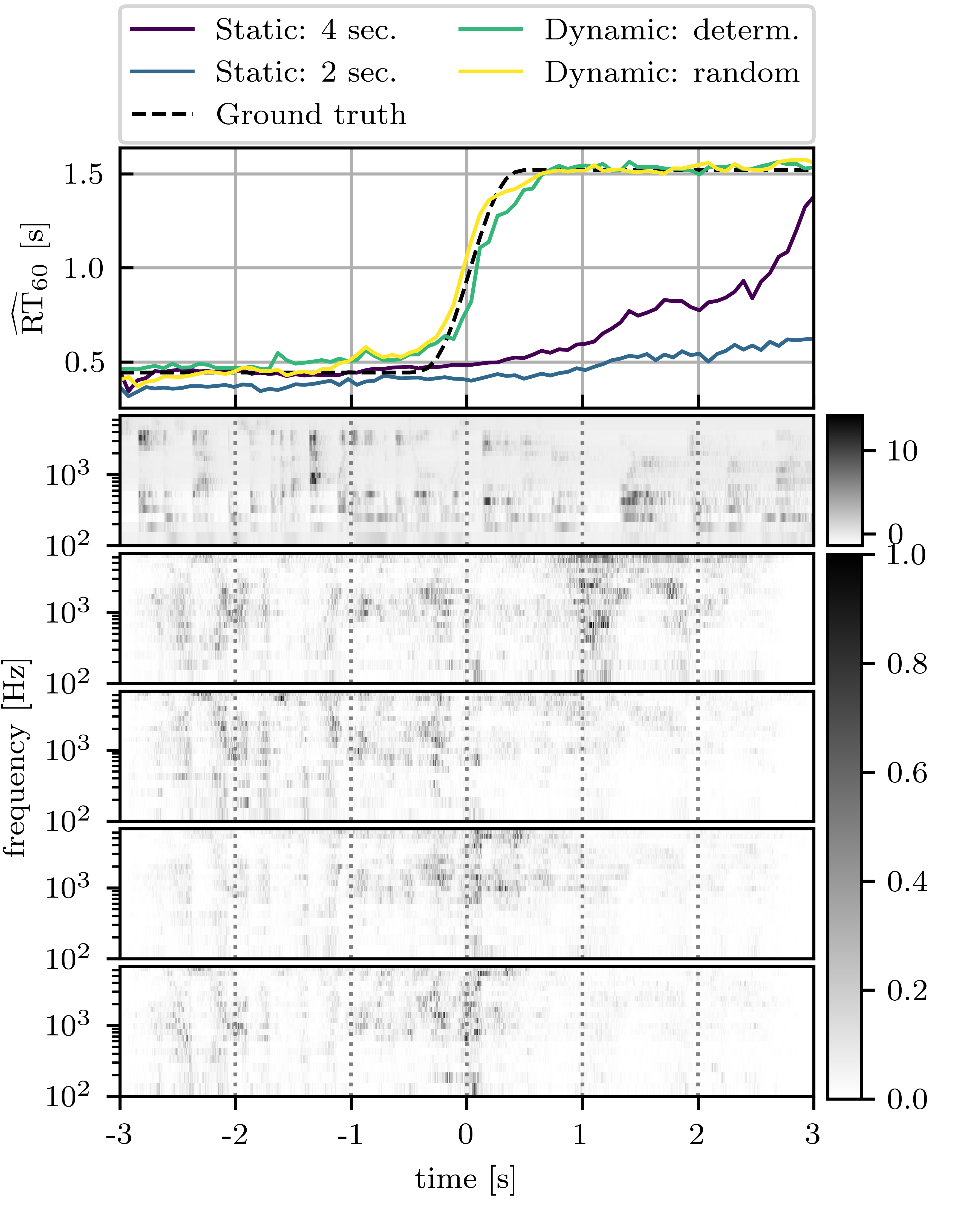}};
    \end{pgfonlayer}
    \begin{pgfonlayer}{annotlayer}
        \node [align=right] (0) at (-1.3,0.4) {\color{black}\scriptsize{Gammatone spectrogram}};
        \node [align=right] (0) at (-1.20,-0.825) {\color{black}\scriptsize{Static training ($4$ sec.)}};
        \node [align=right] (0) at (-1.20,-2.05) {\color{black}\scriptsize{Static training ($2$ sec.)}};
        \node [align=right] (0) at (-1.48,-3.25) {\color{black}\scriptsize{Dynamic training (determ.)}};
        \node [align=right] (0) at (-1.48,-4.5) {\color{black}\scriptsize{Dynamic training (random)}};
    \end{pgfonlayer}
\end{tikzpicture}
\caption{The top plot shows the estimation of the four differently trained networks on a single sample. The input gammatone spectrogram and a comparison of four saliency maps \cite{simonyan2013deep} resulting from the four differently trained models are shown below. \tsixty changes from $0.44\,\mathrm{s}$ in the first half of the sample to $1.52\,\mathrm{s}$ in the second half, the switch occurs at $t = 0\,\mathrm{s}$.}
\label{fig:single_example}
\end{figure}
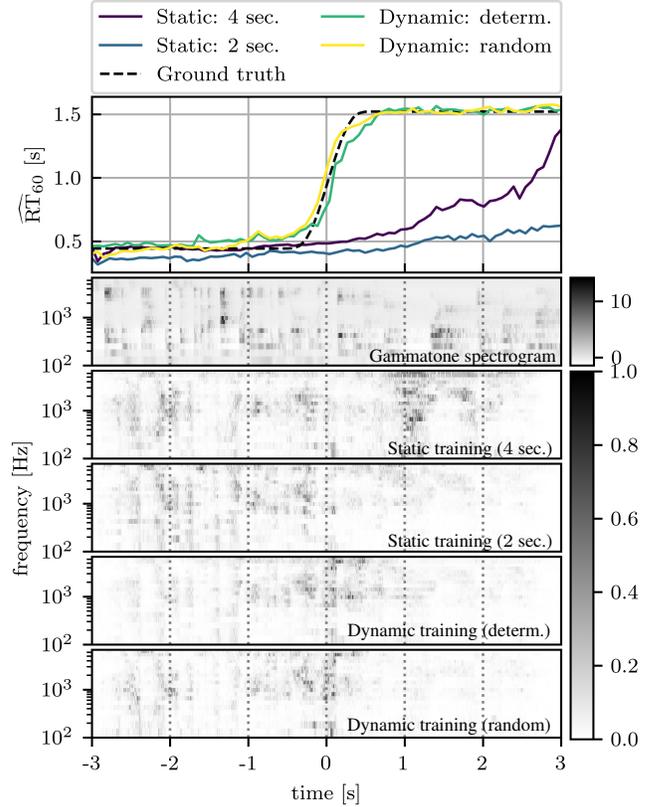

An example is shown in Fig.~\ref{fig:single_example} for a single instance to illustrate the temporal behavior of the model resulting from the four different training methods.
It is observed that the statically trained models cannot track the transition accurately and only react slowly to the changed acoustic conditions, with the statically trained model using two-second segments exhibiting the most stagnant update. The temporal interpolation of the ground truth \tsixty on the output time scale, mentioned in Sec.~\ref{ssec:static_dynamic_data}, can also be seen for this example.

\subsection{Evaluation data}
\label{ssec:eval_data}
Once all models are trained with the data generated according to the four conditions, an evaluation is performed on two separate test sets. For one, a dynamic data set is generated that contains reverberant speech segments with a duration of six seconds, including a change in acoustic conditions after three seconds. This presents all of the differently trained models with dynamic data of unseen length while the fixed change in acoustic conditions after three seconds enables convenient averaging across samples. Additionally, a second test data set representing static acoustic conditions with a duration of ten seconds is generated to investigate if the increased temporal adaptability resulting from dynamic training comes at the cost of reduced accuracy under static conditions. Both test sets use all available source data by generating unique pairings of anechoic speech and AIRs for the four individual training sets.

\section{Performance Evaluation}
\label{sec:eval}
\subsection{Metrics}
\label{ssec:eval_metrics}
\begin{table*}[h!]
    \centering
    \begin{tabular}{r  @{\extracolsep{5pt}} c c @{\extracolsep{5pt}} c c c c @{\extracolsep{5pt}} c c c c}
    \toprule
    & \multicolumn{6}{c}{\textsc{Dynamic test data}} & \multicolumn{4}{c}{\textsc{Static test data}} \\
    \cmidrule{2-7} \cmidrule{8-11} 
    & $\mu$ & $\sigma$ & MSE & $\rho$ & Bias & MAPE & MSE & $\rho$ & Bias & MAPE \\
    \cmidrule{1-7} \cmidrule{8-11} 
    Static training ($4$ sec.) & $0.2185$ & $0.1386$ & $0.1216$ & $0.4881$ & $0.0815$ & $50.89$ & $\mbf{0.0219}$ & $\mbf{0.9308}$ & $-0.0369$ & $18.86$ \\
    Static training ($2$ sec). & $0.2225$ & $0.1457$ & $0.1502$ & $0.4948$ & $0.1922$ & $68.08$ & $0.0347$ & $0.8767$ & $0.044$ & $23.57$ \\
    Dynamic training (determ.) & $0.1773$ & $0.1153$ & $0.0637$ & $0.7677$ & $-0.0632$ & $26.87$ & $0.0278$ & $0.8993$ & $\mbf{-0.0197}$ & $19.88$ \\
    Dynamic training (random) & $\mbf{0.1551}$ & $\mbf{0.1018}$ & $\mbf{0.0509}$ & $\mbf{0.8136}$ & $\mbf{-0.0406}$ & $\mbf{23.86}$ & $0.0268$ & $0.9069$ & $-0.0365$ & $\mbf{18.69}$ \\
    \bottomrule
    \end{tabular}
    \caption{Overview over various metrics describing the performance of the model trained with four different data sets, evaluated on dynamic and static test data.}
    \label{tab:eval}
\end{table*}

Table \ref{tab:eval} provides an overview of various metrics that reflect the performance of the differently trained models under dynamic and static acoustic conditions. In the dynamic case, mean $\mu$ and standard deviation $\sigma$ of the squared error at all time steps in each sample are averaged over the entire test set and are indicators of the overall estimation accuracy and temporal reactivity of the model. Another aspect is the estimation accuracy at the last time step in each sample, after the change in acoustic conditions has occurred and the model processed the entire sample. For both static and dynamic conditions, the last step's performance is evaluated based on four different metrics: MSE, Pearson correlation coefficient $\rho$, bias and mean absolute percentage error (MAPE).

Under dynamic conditions, the two networks trained with dynamic samples exhibit consistently better metrics with a reduction in MSE by over $50\%$ and an improvement of the correlation coefficient by approximately $65\%$ when compared to the static training using four seconds. At the same time, mean and standard deviation are also reduced by approximately $30\%$. Under static acoustic conditions, the dynamically trained models perform only slightly less accurately.

\subsection{Temporal evaluation}
\label{ssec:eval_time}
In addition to the analysis of performance metrics, three aspects that describe the temporal behaviour of the model are investigated qualitatively. The temporal characteristics of the estimation error response to changing acoustic conditions, the degree of renewed network activation based on frequency-averaged saliency maps \cite{simonyan2013deep} and the importance attribution of new information following the acoustic change based on frequency-averaged integrated gradients \cite{sundararajan2017axiomatic}.
\begin{figure}[!t]
    \centering
    \includegraphics[width=\columnwidth,trim={0.0cm 0.1cm 0.0cm 0.0cm}]{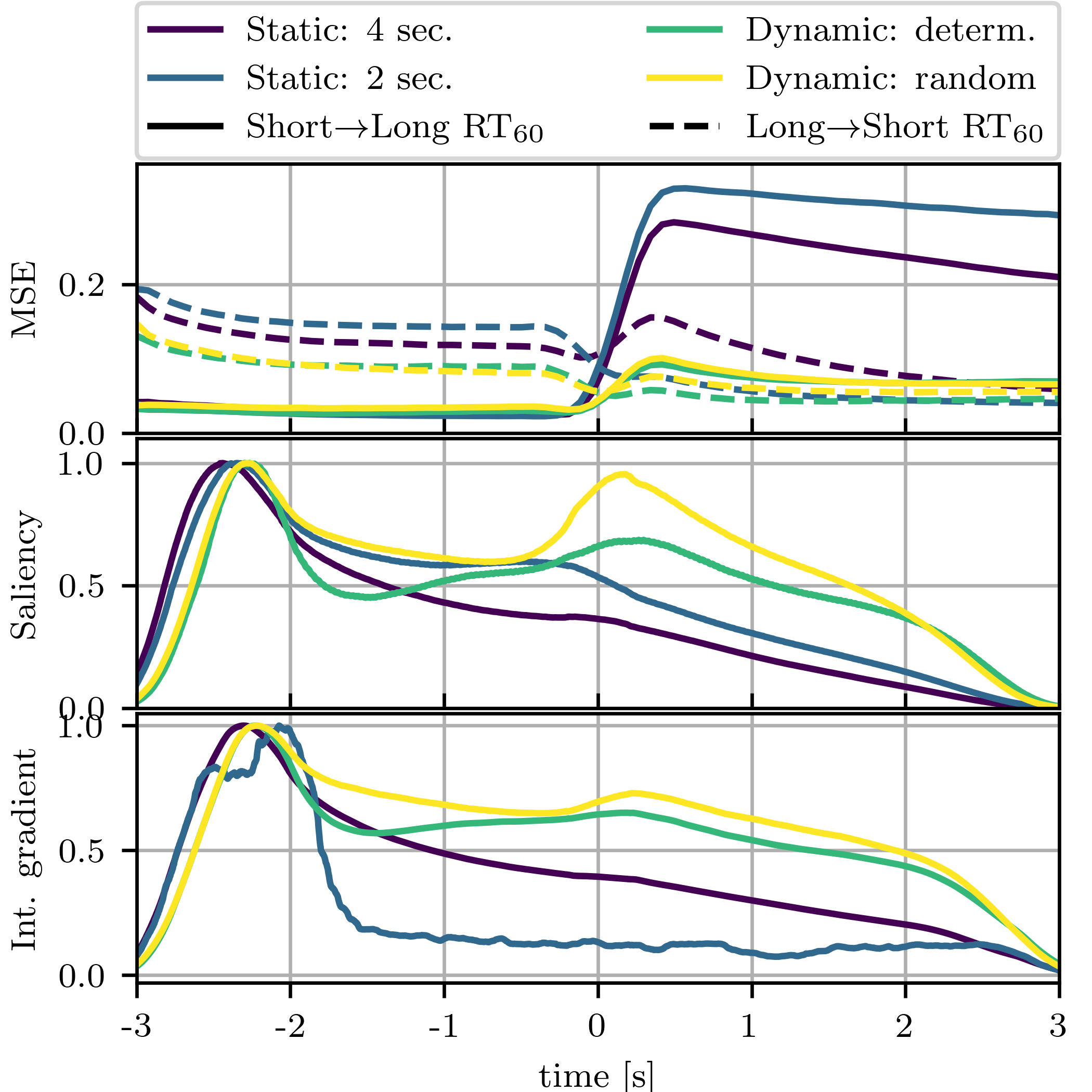}
    \caption{Three plots describing the temporal behaviour of the four trained networks, the change in acoustic conditions occurs at $t=0$. Top: Average squared error separated by the direction of change in \tsixty; Middle: frequency-averaged saliency; Bottom: feature importance attribution based on frequency-averaged integrated gradients.}
    \label{fig:av_all}
\end{figure}
Fig.~\ref{fig:av_all}, top, shows the MSE response for the four training conditions, averaged over the entire test data set and separated by the direction of change in \tsixty. The difference in temporal behaviour is noticeable: when a transition from dry to reverberant acoustic conditions occurs, the statically trained models are only able to gradually update their estimation, while the dynamically trained models are quick to react. Furthermore, the dynamically trained models show improved accuracy with a lower average MSE before the change in conditions for reverberant environments.

Another aspect is the magnitude of the frequency averaged, normalized gradients resulting from the four differently trained networks. The statically trained models do not significantly react to the change in conditions and can only update the estimation gradually. The dynamically trained network's ability to track the acoustic change is indicated by the renewed activation of neurons concentrated around the moment of transition in Fig.~\ref{fig:single_example}, with the \emph{dynamic (random)} training condition exhibiting the largest gradients. Reflected by the average integrated gradients in Fig.~\ref{fig:av_all}, the two statically trained models only attribute importance to the beginning of each sample while later parts are taken into account to a lesser degree. In contrast, the models trained with dynamic samples continuously consider new information with a slight emphasis on the moment of changing acoustic conditions.

One interesting insight arises from the comparison of training conditions \emph{Static ($2$ sec.)} and \emph{Dynamic (deterministic)}. While the total number of pairings of anechoic speech and AIRs seen by the model during training is equal in both conditions, the dynamically trained model performs considerably more accurately when acoustic conditions change. This observation supports the hypothesis that a portion of the convolutional features and recurrent modeling learned by the dynamically trained network relates to the detection of and reaction to changing acoustic conditions.

\section{Conclusion}
\label{sec:conclusion}
We investigated a previously proposed neural network \cite{deng2020online} for blind \tsixty estimation under dynamic acoustic conditions and identified an insufficient ability to track temporal changes in the acoustic properties of the environment. We proposed a novel way of generating training data that results in a considerable improvement in the model's capability of estimating changing \tsixty while incurring only a slight loss of estimation accuracy under static acoustic conditions. The findings presented in this study illustrate the great potential of data-centric approaches to existing problems in acoustic scene analysis in particular and machine learning in general.


\bibliographystyle{IEEEbib}
{\footnotesize\bibliography{main}}

\begin{thebibliography}{10}

\bibitem{iso3382}
``{ISO} 3382-1:2009({EN}): Acoustics — measurement of room acoustic
  parameters,'' 2009.

\bibitem{stowell2015detection}
Dan Stowell, Dimitrios Giannoulis, Emmanouil Benetos, Mathieu Lagrange, and
  Mark~D. Plumbley,
\newblock ``Detection and classification of acoustic scenes and events,''
\newblock {\em IEEE Transactions on Multimedia}, vol. 17, no. 10, pp.
  1733--1746, 2015.

\bibitem{benesty2011speech}
Jacob Benesty, Jingdong Chen, and Emanu{\"e}l~A.P. Habets,
\newblock {\em Speech enhancement in the STFT domain},
\newblock Springer Science \& Business Media, 2011.

\bibitem{remaggi2019perceived}
Luca Remaggi, Kim Hansung, Annika Neidhardt, Adrian Hilton, and J.B. Philip,
\newblock ``Perceived quality and spatial impression of room reverberation in
  {VR} reproduction from measured images and acoustics,''
\newblock in {\em Proc. of the 23rd Int. Congr. Acoust}, 2019.

\bibitem{eaton2016estimation}
James Eaton, Nikolay~D. Gaubitch, Alastair~H. Moore, and Patrick~A. Naylor,
\newblock ``Estimation of room acoustic parameters: The {ACE} challenge,''
\newblock {\em {IEEE} Trans. Audio, Speech, Lang. Process.}, vol. 24, no. 10,
  pp. 1681--1693, 2016.

\bibitem{malik2010audio}
Hafiz Malik and Hany Farid,
\newblock ``Audio forensics from acoustic reverberation,''
\newblock in {\em Proc. {IEEE} Intl. Conf. on Acoustics, Speech and Signal
  Processing (ICASSP)}, 2010, pp. 1710--1713.

\bibitem{ratnam2003blind}
Rama Ratnam, Douglas~L. Jones, Bruce~C. Wheeler, William~D. O’Brien~Jr.,
  Charissa~R. Lansing, and Albert~S. Feng,
\newblock ``Blind estimation of reverberation time,''
\newblock {\em The Journal Acoust. Soc. of America}, vol. 114, no. 5, pp.
  2877--2892, 2003.

\bibitem{lollmann2008estimation}
Heinrich~W. L{\"o}llmann and Peter Vary,
\newblock ``Estimation of the reverberation time in noisy environments,''
\newblock in {\em Proc. Intl. Workshop Acoust. Signal Enhancement ({IWAENC})},
  2008.

\bibitem{6637629}
James Eaton, Nikolay~D. Gaubitch, and Patrick~A. Naylor,
\newblock ``Noise-robust reverberation time estimation using spectral decay
  distributions with reduced computational cost,''
\newblock in {\em Proc. {IEEE} Intl. Conf. on Acoustics, Speech and Signal
  Processing (ICASSP)}, 2013, pp. 161--165.

\bibitem{lollmann2010improved}
Heiner L{\"o}llmann, Emre Yilmaz, Marco Jeub, and Peter Vary,
\newblock ``An improved algorithm for blind reverberation time estimation,''
\newblock in {\em Proc. Intl. Workshop Acoust. Signal Enhancement ({IWAENC})},
  2010, pp. 1--4.

\bibitem{wen2008blind}
Jimi~Y.C. Wen, Emanu{\"e}l~A.P. Habets, and Patrick~A. Naylor,
\newblock ``Blind estimation of reverberation time based on the distribution of
  signal decay rates,''
\newblock in {\em Proc. {IEEE} Intl. Conf. on Acoustics, Speech and Signal
  Processing (ICASSP)}, 2008, pp. 329--332.

\bibitem{gamper2018blind}
Hannes Gamper and Ivan~J. Tashev,
\newblock ``Blind reverberation time estimation using a convolutional neural
  network,''
\newblock in {\em Proc. Intl. Workshop Acoust. Signal Enhancement ({IWAENC})},
  2018, pp. 136--140.

\bibitem{duangpummet2021blind}
Suradej Duangpummet, Jessada Karnjana, Waree Kongprawechnon, and Masashi Unoki,
\newblock ``Blind estimation of room acoustic parameters and speech
  transmission index using {MTF}-based {CNNs},''
\newblock {\em arXiv preprint arXiv:2103.07904}, 2021.

\bibitem{callens2020joint}
Paul Callens and Milos Cernak,
\newblock ``Joint blind room acoustic characterization from speech and music
  signals using convolutional recurrent neural networks,''
\newblock {\em arXiv preprint arXiv:2010.11167}, 2020.

\bibitem{perez2019machine}
Ricardo~Falcon Perez, Georg G{\"o}tz, and Ville Pulkki,
\newblock ``Machine-learning-based estimation of reverberation time using room
  geometry for room effect rendering,''
\newblock in {\em Proc. of the International Congress on Acoustics: integrating
  4th EAA Euroregio}, 2019, vol.~9, p.~13.

\bibitem{deng2020online}
Shuwen Deng, Wolfgang Mack, and Emanu{\"e}l~A.P. Habets,
\newblock ``Online blind reverberation time estimation using {CRNNs},''
\newblock in {\em Proc. Interspeech Conf.}, 2020, pp. 5061--5065.

\bibitem{sarroff2020blind}
Andy Sarroff and Roth Michaels,
\newblock ``Blind arbitrary reverb matching,''
\newblock in {\em Proc. Conf. on Digital Audio Effects}, 2020.

\bibitem{steinmetz2021filtered}
Christian~J. Steinmetz, Vamsi~Krishna Ithapu, and Paul Calamia,
\newblock ``Filtered noise shaping for time domain room impulse response
  estimation from reverberant speech,''
\newblock {\em arXiv preprint arXiv:2107.07503}, 2021.

\bibitem{patterson1987efficient}
Roy~D. Patterson, Ian Nimmo-Smith, John Holdsworth, and Peter Rice,
\newblock ``An efficient auditory filterbank based on the gammatone function,''
\newblock in {\em A meeting of the IOC Speech Group on Auditory Modelling at
  RSRE}, 1987, vol.~2.

\bibitem{gold2011speech}
Ben Gold, Nelson Morgan, and Dan Ellis,
\newblock {\em Speech and audio signal processing: processing and perception of
  speech and music},
\newblock John Wiley \& Sons, 2011.

\bibitem{akiba2019optuna}
Takuya Akiba, Shotaro Sano, Toshihiko Yanase, Takeru Ohta, and Masanori Koyama,
\newblock ``Optuna: A next-generation hyperparameter optimization framework,''
\newblock in {\em Proc. of the 25th ACM SIGKDD international conference on
  knowledge discovery \& data mining}, 2019, pp. 2623--2631.

\bibitem{araujo2019computing}
Andr{\'e} Araujo, Wade Norris, and Jack Sim,
\newblock ``Computing receptive fields of convolutional neural networks,''
\newblock {\em Distill}, vol. 4, no. 11, pp. e21, 2019.

\bibitem{lecun2012efficient}
Yann~A. LeCun, L{\'e}on Bottou, Genevieve~B. Orr, and Klaus-Robert M{\"u}ller,
\newblock ``Efficient backprop,''
\newblock in {\em Neural networks: Tricks of the trade}, pp. 9--48. Springer,
  2012.

\bibitem{antsalo2001estimation}
Poju Antsalo, Aki Makivirta, Vesa Valimaki, Timo Peltonen, and Matti
  Karjalainen,
\newblock ``Estimation of modal decay parameters from noisy response
  measurements,''
\newblock in {\em Audio Engineering Society Convention 110}, 2001.

\bibitem{panayotov2015librispeech}
Vassil Panayotov, Guoguo Chen, Daniel Povey, and Sanjeev Khudanpur,
\newblock ``Librispeech: an {ASR} corpus based on public domain audio books,''
\newblock in {\em Proc. {IEEE} Intl. Conf. on Acoustics, Speech and Signal
  Processing (ICASSP)}, 2015, pp. 5206--5210.

\bibitem{zue1990speech}
Victor Zue, Stephanie Seneff, and James Glass,
\newblock ``Speech database development at {MIT}: {TIMIT} and beyond,''
\newblock {\em Speech Communication}, vol. 9, no. 4, pp. 351--356, 1990.

\bibitem{schroder2011open}
Marc Schr{\"o}der, Marcela Charfuelan, Sathish Pammi, and Ingmar Steiner,
\newblock ``Open source voice creation toolkit for the {MARY} {TTS}
  {Platform},''
\newblock in {\em Annual Conference of the International Speech Communication
  Association}, 2011.

\bibitem{jeub2009binaural}
Marco Jeub, Magnus Schafer, and Peter Vary,
\newblock ``A binaural room impulse response database for the evaluation of
  dereverberation algorithms,''
\newblock in {\em 2009 16th International Conference on Digital Signal
  Processing}, 2009, pp. 1--5.

\bibitem{murphy2010openair}
Damian~T. Murphy and Simon Shelley,
\newblock ``Openair: An interactive auralization web resource and database,''
\newblock in {\em Audio Engineering Society Convention 129}, 2010.

\bibitem{echothief}
``Echothief impulse response library,'' http://www.echothief.com/,
\newblock Accessed: 2021-10-05.

\bibitem{allen1979image}
Jont~B. Allen and David~A. Berkley,
\newblock ``Image method for efficiently simulating small-room acoustics,''
\newblock {\em The Journal Acoust. Soc. of America}, vol. 65, no. 4, pp.
  943--950, 1979.

\bibitem{tieleman2012lecture}
Tijmen Tieleman, Geoffrey Hinton, et~al.,
\newblock ``Lecture 6.5-rmsprop: Divide the gradient by a running average of
  its recent magnitude,''
\newblock {\em COURSERA: Neural networks for machine learning}, vol. 4, no. 2,
  pp. 26--31, 2012.

\bibitem{simonyan2013deep}
Karen Simonyan, Andrea Vedaldi, and Andrew Zisserman,
\newblock ``Deep inside convolutional networks: Visualising image
  classification models and saliency maps,''
\newblock {\em arXiv preprint arXiv:1312.6034}, 2013.

\bibitem{sundararajan2017axiomatic}
Mukund Sundararajan, Ankur Taly, and Qiqi Yan,
\newblock ``Axiomatic attribution for deep networks,''
\newblock in {\em Proc. Intl. Conf. Machine Learning ({ICML})}, 2017, pp.
  3319--3328.

\end{thebibliography}

\end{document}